\documentclass[10pt,a4paper,conference]{IEEEtran_icc}
\usepackage{amsmath,amssymb,amsthm}
\usepackage{graphicx,color,enumerate}
\usepackage{algpseudocode}
\usepackage{algorithm}

\newtheorem{mydef}{Definition}

\newcommand{\fis}[3]{\frac{1+\sum_{#2\neq #1}|h_{#1#2}|^2 #3_{#2}(h)}{|h_{#1#1}|^2}}
\newcommand{\fij}[3]{\frac{1+\sum_{#2}|h_{#1#2}|^2 #3_{#2}(h)}{|h_{#1#1}|^2}}

\title{Power Allocation Games on Interference Channels with Complete and Partial Information}
\author{\IEEEauthorblockN{Krishna Chaitanya A, Utpal Mukherji, Vinod Sharma}
\IEEEauthorblockA{Department of ECE, Indian Institute of Science, Bangalore-560012 \\ Email: $\lbrace$akc, utpal, vinod$\rbrace$ @ece.iisc.ernet.in}
}
\date{}
\begin{document}
\maketitle
\begin{abstract}
We consider a wireless channel shared by multiple transmitter-receiver pairs. Their transmissions interfere with each other.  Each transmitter-receiver pair aims to maximize its long-term average transmission rate subject to an average power constraint.  This scenario is modeled as a stochastic game under different assumptions.  We first assume that each transmitter and receiver has knowledge of all direct and cross link channel gains.  We later relax the assumption to the knowledge of incident channel gains and then further relax to the knowledge of the direct link channel gains only.  In all the cases, we formulate the problem of finding the Nash equilibrium as a variational inequality (VI) problem and present an algorithm to solve the VI.
\end{abstract}
\section{Introduction}
\hspace{0.5cm} We consider a wireless channel which is being shared by multiple users to transmit their data to their respective receivers.  The transmissions of different users may cause interference to other receivers.  This is a typical scenario in many wireless networks.  In particular, this can represent inter-cell interference on a particular wireless channel in a cellular network.  The different users want to maximize their transmission rates.  This system can be modeled in the game theoretic framework and has been widely studied \cite{palomar} - \cite{async}.\par
\hspace{0.5cm} In \cite{palomar}, the authors have considered parallel Gaussian interference channels.  This setup is modeled as a strategic form game and existence and uniqueness of a Nash equilibrium (NE) is studied.  The authors provide conditions under which the water-filling function is a contraction and thus obtain conditions for uniqueness of NE and for convergence of iterative water-filling.  They extend these results to a multi-antenna system in \cite{MIMO_IWF} and consider an asynchronous version of iterative water-filling in \cite{async}. \par
\hspace{0.5cm} An online algorithm to reach a NE for the parallel Gaussian channels is presented in \cite{stochastic} when the channel gain distributions are not known to the players.  Its convergence is also proved.  In \cite{VI} authors describe some conditions under which parallel Gaussian interference channels have multiple Nash equilibria.  Using variational inequalities, they present an algorithm that converges to a Nash equilibrium which minimizes the overall weighted interference.\par
\hspace{0.5cm}    We consider power allocation in a non-game-theoretic framework in \cite{PA} (see also other references in \cite{PA} for such a setup).  In \cite{PA}, we have proposed a centralized algorithm for finding the Pareto points that maximize sum rate when the receivers have knowledge of all the channel gains and decode the messages from strong and very strong interferers instead of treating them as noise.\par
\hspace{0.5cm} All the above cited works consider a one shot non-cooperative game (or a Pareto point).  As against that we consider a stochastic game over Gaussian interference channels, where the users want to maximize their long term average rate and have long term average power constraints (for potential advantages of this over one shot optimization, see \cite{goldsmith}, \cite{vsharma}).  For this system we obtain existence of NE and develop algorithms to obtain NE via variational inequalities.  Further more, the above mentioned literature considers the problem when each user knows all the channel gains in the system while we also consider the much more realistic situation when a user knows only its own channel gains.\par
\hspace{0.5cm} The paper is organized as follows. In Section \ref{sys_model}, we present the system model and formulate it as a stochastic game.  In Section \ref{one}, we study this stochastic game and define the basic terminology.  In Section \ref{gen_vi}, we propose an algorithm to solve the formulated variational inequality under general conditions.  In Section \ref{incomplete} we use this algorithm to obtain NE when the users have only partial information about the channel gains.  In Section \ref{ne}, we present numerical examples and Section \ref{concl} concludes the paper.
\section{System model and Notation} \label{sys_model}
\hspace{0.5cm} We consider a Gaussian wireless channel being shared by $N$ transmitter-receiver pairs. The time axis is slotted and all users' slots are synchronized.  The channel gains of each transmit-receive pair are constant during a slot and change independently from slot to slot.  These assumptions are usually made for this system \cite{palomar}, \cite{vsharma}.\par
\hspace{0.5cm} Let $H_{ij}(k)$ be the random variable that represents channel gain from transmitter $j$ to receiver $i$ (for transmitter $i$, receiver $i$ is the intended receiver) in slot $k$.  The direct channel power gains $\vert H_{ii}(k)\vert^2 \in \mathcal{H}_d = \lbrace g_1^{(d)},g_2^{(d)},\dots,g_{n_1}^{(d)} \rbrace$ and the cross channel power gains $\vert H_{ij}(k)\vert^2 \in \mathcal{H}_c = \lbrace g_1^{(c)},g_2^{(c)},\dots,g_{n_2}^{(c)}\rbrace$.  Let $\pi_d$ and $\pi_c$ be the probability distributions on $\mathcal{H}_d$ and $\mathcal{H}_c$ respectively.  We assume that, $\{H_{ij}(k), k \geq 0 \}$ is an $i.i.d$ sequence with distribution $\pi_{ij}$ where $\pi_{ij} = \pi_d$ if $i=j$ and $\pi_{ij} = \pi_c$ if $i\neq j$.  We also assume that these sequences are independent of each other.\par
\hspace{0.5cm} We denote $(H_{ij}(k), i,j = 1,\dots,N)$ by ${\bf H}(k)$ and its realization vector by $h(k)$ which takes values in $\mathcal{H}$, the set of all possible channel states.  The distribution of ${\bf H}(k)$ is denoted by $\pi$.  We call the channel gains $(H_{ij}(k), j = 1,\dots,N)$ from all the transmitters to the receiver $i$ an incident gain of user $i$ and denote by ${\bf H}_i(k)$ and its realization vector by $h_i(k)$ which takes values in $\mathcal{I}$, the set of all possible incident channel gains.  The distribution of ${\bf H}_i(k)$ is denoted by $\pi_I$.\par
\hspace{0.5cm} Each user aims to operate at a power allocation that maximizes its long term average rate under an average power constraint.  Since their transmissions interfere with each other, affecting their transmission rates,  we model this scenario as a stochastic game.\par
\hspace{0.5cm} We first assume complete channel knowledge at all transmitters and receivers.  If user $i$ uses power $P_i({\bf H}(k))$ in slot $k$, it gets rate $\text{log} \left(1+\Gamma_i \left(P \left({\bf H}(k) \right) \right) \right)$, where
\begin{equation}
\Gamma_i(P({\bf H}(k))) = \frac{\alpha_i |H_{ii}(k)|^2 P_i({\bf H}(k))}{1 + \sum_{j \neq i}|H_{ij}(k)|^2P_j({\bf H}(k))},
\end{equation}
$P({\bf H}(k)) = (P_1({\bf H}(k)),\dots,P_N({\bf H}(k)))$ and $\alpha_i$ is a constant that depends on the modulation and coding used by transmitter $i$ and we assume $\alpha_i = 1$ for all $i$.  The aim of each user $i$  is to choose a power policy to maximize its long term average rate
\begin{equation}
r_i({\bf P}_i,{\bf P}_{-i}) \triangleq \limsup\limits_{n \rightarrow \infty} \frac{1}{n} \sum_{k=1}^n \mathbb{E}[\text{log} \left(1+\Gamma_i \left(P \left({\bf H}(k) \right) \right) \right)],
\end{equation}
subject to average power constraint
\begin{equation}
\limsup\limits_{n \rightarrow \infty} \frac{1}{n} \sum_{k=1}^n \mathbb{E}[P_i({\bf H}(k))] \leq \overline{P}_i, \text{ for each } i,\label{avg_c}
\end{equation}
where ${\bf P}_{-i}$ denotes the power policies of all users except $i$.  We denote this game by $\mathcal{G}_A$.\par
\hspace{0.5cm} We next assume that the $i$th transmitter-receiver pair has knowledge of its incident gains ${\bf H}_i$ only.  Then the rate of user $i$ is
\begin{multline}
r_i({\bf P}_i,{\bf P}_{-i}) \triangleq \\ \limsup\limits_{n \rightarrow \infty} \frac{1}{n} \sum_{k=1}^n \mathbb{E}_{{\bf H}_i(k)} \left[\mathbb{E}_{{\bf H}_{-i}(k)}[\text{log} \left(1+ \Gamma_i({\bf H}_i(k),{\bf H}_{-i}(k))\right)]\right],
\end{multline}
where $P_i({\bf H}(k))$ depends only on ${\bf H}_i(k)$ and $\mathbb{E}_X$ denotes expectation with respect to the distribution of $X$.  Each user maximizes its rate subject to (\ref{avg_c}), we denote this game by $\mathcal{G}_I$.\par
\hspace{0.5cm} We also consider a game assuming that each transmitter-receiver pair knows only its direct link gain $H_{ii}$.  This is the most realistic assumption since each receiver $i$ can estimate $H_{ii}$ and feed it back to transmitter $i$.  In this case, the rate of user $i$ is given by
\begin{multline}
r_i({\bf P}_i,{\bf P}_{-i}) \triangleq \limsup\limits_{n \rightarrow \infty} \frac{1}{n} \sum_{k=1}^n \mathbb{E}_{{\bf H}_{ii}(k)} \left[ \mathbb{E}_{{\bf H}_{-ii}(k)}\right.\\ \left. [\text{log} \left(1+ \Gamma_i(H_{ii}(k),H_{-ii}(k))\right)]\right], \label{r_d}
\end{multline}
where $P_i({\bf H}(k))$ is a function of $H_{ii}(k)$ only.  Here, $H_{-ii}$ denotes the channel gains of all other links in the interference channel except $H_{ii}$.  In this game, each user maximizes its rate (\ref{r_d}) under the average power constraint (\ref{avg_c}).  We denote this game by $\mathcal{G}_D$.\par
\hspace{0.5cm} We address these problems as stochastic games with the set of feasible power policies of user $i$ denoted by $\mathcal{A}_i$ and its utility by $r_i$.  Let $\mathcal{A} = \Pi_{i=1}^{N} \mathcal{A}_i$.\par
\hspace{0.5cm} We limit ourselves to stationary policies, i.e., the power policy for every user in slot $k$ depends only on the channel state $H(k)$ and not on $k$.  In the current setup, it does not entail any loss in optimality.  In fact now we can rewrite the optimization problem in $\mathcal{G}$ to find policy $P({\bf H})$ such that $r_i = \mathbb{E}_{\bf H}[\text{log} \left(1+\Gamma_i \left(P \left({\bf H}\right) \right) \right)]$ is maximized subject to $\mathbb{E}_{\bf H}\left[P_i({\bf H})\right] \leq \overline{P}_i$ for all $i$.  Similarly, we can rewrite the optimization problems in games $\mathcal{G}_I$ and $\mathcal{G}_D$.  We express power policy of player $i$ by ${\bf P}_i = (P_i(h), h\in \mathcal{H})$, where transmitter $i$ transmits in channel state $h$ with power $P_i(h)$.  We denote the power profile of all players by ${\bf P} = ({\bf P}_1,\dots,{\bf P}_N)$. \par
\hspace{0.5cm} In the rest of the paper, we prove existence of a Nash equilibrium for each of these games and provide algorithm to compute it.\par
\section{Game Theoretic Reformulation} \label{one}
\hspace{0.5cm} Theory of variational inequalities offers various algorithms to find NE of a given game \cite{Pang}.  A variational inequality problem denoted by $VI(K,F)$ is defined as follows.
\begin{mydef}
Let $K \subset \mathbb{R}^n$ be a closed and convex set, and $F:K \to K$. The variational inequality problem $VI(K,F)$ is defined as the problem of finding $x \in K$ such that $$F(x)^T(y-x) \geq 0 \text{ for all } y \in K.$$
\end{mydef}
\hspace{0.5cm} We reformulate the Nash equilibrium problem at hand to an affine variational inequality problem.  We denote our game by $\mathcal{G} = \big((\mathcal{A}_i)_{i=1}^N, (r_i)_{i=1}^N\big)$, where $r_i({\bf P}_i,{\bf P}_{-i}) = \mathbb{E}_{\bf H}[\text{log} \left(1+\Gamma_i \left(P \left({\bf H}\right) \right) \right)]$ and $\mathcal{A}_i = \{{\bf P}_i \in \mathbb{R}^N: \mathbb{E}_{\bf H}\left[P_i({\bf H})\right] \leq \overline{P}_i, P_i(h) \geq 0 \text{ for all } h \in \mathcal{H}\}$.
\begin{mydef}
A point ${\bf P}^*$ is a Nash Equilibrium (NE) of game $\mathcal{G} = \big((\mathcal{A}_i)_{i=1}^N, (r_i)_{i=1}^N\big)$ if for each player $i$
\begin{equation*}
r_i({\bf P}_i^*,{\bf P}_{-i}^*) \geq r_i({\bf P}_i,{\bf P}_{-i}^*) \text{ for all } {\bf P}_i \in \mathcal{A}_i.
\end{equation*}
\end{mydef}
\hspace{0.5cm} Existence of a pure NE for the strategic games $\mathcal{G}_A, \mathcal{G}_I$ and $\mathcal{G}_D$ follows from the Debreu-Glicksberg-Fan Theorem (\cite{BASAR}, page no. 69), since in our game $r_i({\bf P}_i,{\bf P}_{-i})$ is a continuous function in the profile of strategies ${\bf P} = ({\bf P}_i,{\bf P}_{-i}) \in \mathcal{A}$ and concave in ${\bf P}_i$ for $\mathcal{G}_A, \mathcal{G}_I$ and $\mathcal{G}_D$.
\begin{mydef}
  The best-response of player $i$ is a function $BR_i : \mathcal{A}_{-i} \rightarrow \mathcal{A}_i$ such that $BR_i({\bf P}_{-i})$ maximizes $r_i({\bf P}_i, {\bf P}_{-i})$, subject to ${\bf P}_i \in \mathcal{A}_i$.   
\end{mydef}
\hspace{0.5cm}  We see that the Nash equilibrium is a fixed point of the best-response function.  In the following we provide algorithms to obtain this fixed point for $\mathcal{G}_A$.  In Section \ref{incomplete} we will consider $\mathcal{G}_I$ and $\mathcal{G}_D$.  Given other players' power profile ${\bf P}_{-i}$, we use Lagrange method to evaluate the best response of player $i$.  The Lagrangian function is defined by
\begin{equation*}
\mathcal{L}_i({\bf P}_i,{\bf P}_{-i}) = r_i({\bf P}_i, {\bf P}_{-i}) + \lambda_i( \overline{P}_i - \mathbb{E}_{\bf H}\left[P_i({\bf H})\right]).
\end{equation*}
To maximize $\mathcal{L}_i({\bf P}_i,{\bf P}_{-i})$, we solve for ${\bf P}_i$ such that $\frac{\partial \mathcal{L}_i}{\partial {\bf P}_i(h)} = 0$ for each $h \in \mathcal{H}$.  Thus, the component of the best response of player $i$, ${\bf BR}_i({\bf P}_{-i})$ corresponding to channel state $h$ is given by
\begin{multline}
BR_i({\bf P}_{-i};h) = \\ \text{max}\left\{0, \lambda_i({\bf P}_{-i}) - \frac{(1+\sum_{j\neq i}\vert h_{ij}\vert^2 P_j(h))}{\vert h_{ii}\vert^2}\right\}, \label{wf}
\end{multline}
where $\lambda_i({\bf P}_{-i})$ is chosen such that the average power constraint is satisfied.\par
\hspace{0.5cm} It is easy to observe that the best-response of player $i$ to a given strategy of other players is water-filling on ${\bf f}_i({\bf P}_{-i}) = (f_i({\bf P}_{-i};h),h \in \mathcal{H})$ where 
\begin{equation}
f_i({\bf P}_{-i};h) = \frac{(1+\sum_{j\neq i}\vert h_{ij}\vert^2 P_j(h))}{\vert h_{ii}\vert^2}.
\end{equation}
For this reason, we represent the best-response of player $i$ by ${\bf WF}_i({\bf P}_{-i})$.  The notation used for the overall best-response ${\bf WF}({\bf P}) = ({\bf WF}(P(h)), h \in \mathcal{H})$, where ${\bf WF}(P(h)) = (WF_1({\bf P}_{-1};h),\dots,WF_N({\bf P}_{-N};h))$ and $WF_i({\bf P}_{-i};h)$ is as defined in (\ref{wf}).  We use ${\bf WF}_i({\bf P}_{-i}) = (WF_i({\bf P}_{-i};h), h \in \mathcal{H})$.\par 
\hspace{0.5cm} It is observed in \cite{palomar} that the best-response ${\bf WF}_i({\bf P}_{-i})$ is also the solution of the optimization problem
\begin{equation}
\text{ minimize } \left\Vert {\bf P}_i  + {\bf f}_i({\bf P}_{-i})\right\Vert^2, \text{ subject to } {\bf P}_i \in \mathcal{A}_i. \label{opt_proj}
\end{equation}
As a result we can interpret the best-response as the projection of $(-f_{i,1}({\bf P}_{-i}),\dots,-f_{i,N}({\bf P}_{-i}))$ on to $\mathcal{A}_i$.  We denote the projection of $x$ on to $\mathcal{A}_i$ by $\Pi_{\mathcal{A}_i}(x)$.  We consider (\ref{opt_proj}), as a game in which every player minimizes its cost function $\left\Vert {\bf P}_i  + {\bf f}_i({\bf P}_{-i})\right\Vert^2$ with strategy set of player $i$ being $\mathcal{A}_i$.  We denote this game by $\mathcal{G}^{\prime}$.  This game has the same set of NEs as $\mathcal{G}$ because the best responses of these two games are equal.  We now formulate the variational inequality problem corresponding to the game $\mathcal{G}^{\prime}$.\par
\hspace{0.5cm} Observe that (\ref{opt_proj}) is a convex optimization problem. Given ${\bf P}_{-i}$, a necessary and sufficient condition for ${\bf P}_i^*$ to be a solution of the convex optimization problem of player $i$ (\cite{Comp}, page 210) is given by
\begin{equation}
\sum_{h \in \mathcal{H}} \left( P_i^*(h) + f_i({\bf P}_{-i};h) \right)\left( x_i(h) - P_{i}^*(h) \right)\geq 0,\label{ineq}
\end{equation}
for all ${\bf x}_{i} \in \mathcal{A}_i$.  Thus, ${\bf P}^{*}$ is a NE of the game $\mathcal{G}^{\prime}$ if (\ref{ineq}) holds
for each player $i$.  We can rewrite the $N$ inequalities in (\ref{ineq}) in compact form as 
\begin{equation}
\left({\bf P}^* + \hat{h} + \hat{H}{\bf P}^* \right)^T \left(x - {\bf P}^*\right) \geq 0 \text{ for all } x \in \mathcal{A}, \label{cond_1}
\end{equation}
where $\hat{h}$ is a $N_1$-length block vector with $N_1 = \vert \mathcal{H}\vert$, the cardinality of $\mathcal{H}$, each block $\hat{h}(h), h \in \mathcal{H}$, is of length $N$ and is defined by $\hat{h}(h) = \left( \frac{1}{\vert h_{11} \vert^2}, \dots, \frac{1}{\vert h_{NN} \vert^2}\right)$ and  $\hat{H}$ is the block diagonal matrix $\hat{H} = \text{diag}\left\lbrace \hat{H}(h), h \in \mathcal{H} \right\rbrace$ with each block $\hat{H}(h)$ defined by
\begin{equation*}
  [\hat{H}(h)]_{ij} = \begin{cases}
    0 & \text{ if } i=j, \\
    \frac{\vert h_{ij} \vert^2}{\vert h_{ii} \vert^2}, & \text{ else. }
    \end{cases}
\end{equation*}
\hspace{0.5cm} The characterization of Nash equilibrium in (\ref{cond_1}) corresponds to solving for ${\bf P}$ in the variational inequality problem $VI(\mathcal{A},F)$,
\begin{equation*}
F({\bf P})^T \left(x - {\bf P}\right) \geq 0 \text{ for all } x \in \mathcal{A},
\end{equation*}
where $F({\bf P}) =  (I+ \hat{H}){\bf P} + \hat{h}$.
\section{Solving the VI for general channels}\label{gen_vi}
\hspace{0.5cm} In \cite{arxiv}, we proved that if $\tilde{H} = (I+ \hat{H})$ is positive semidefinite, then the fixed point iteration 
\begin{equation}
{\bf P}^{(n)} = \Pi_{\mathcal{A}}({\bf P}^{(n-1)}- \tau F({\bf P}^{(n-1)})),\label{f_iter}
\end{equation}
converges to a NE.  This condition is much weaker than one would obtain by using the methods in \cite{palomar}.  In the current setup we aim to find a NE even if $\tilde{H}$ is not positive semidefinite.  For this, we present an algorithm to solve the $VI(\mathcal{A},F)$ in general. \par
\hspace{0.5cm} We note that a solution ${\bf P}^*$ of $VI(\mathcal{A},F)$ satisfies 
\begin{equation}
{\bf P}^* = \Pi_{\mathcal{A}}\left( {\bf P}^* - \tau F({\bf P}^*)\right).
\end{equation}
Thus, ${\bf P}^*$ is a fixed point of the mapping $T({\bf P}) = \Pi_{\mathcal{A}}\left( {\bf P} - F({\bf P})\right)$.  Using this fact, we reformulate the variational inequality problem as a non-convex optimization problem
\begin{eqnarray}
&\text{ minimize } & \Vert {\bf P} - \Pi_{\mathcal{A}}\left( {\bf P} - F({\bf P})\right)\Vert^2, \nonumber \\
&\text{ subject to } & {\bf P} \in \mathcal{A}.\label{s_obj}
\end{eqnarray}
The feasible region $\mathcal{A}$ of ${\bf P}$, can be written as a Cartesian product of $\mathcal{A}_i$, for each $i$, as the constraints of each player are decoupled in power variables.  As a result, we can split the projection $\Pi_{\mathcal{A}}(.)$ into multiple projections $\Pi_{\mathcal{A}_i}(.)$ for each $i$, i.e., $\Pi_{\mathcal{A}}({\bf x}) = (\Pi_{\mathcal{A}_1}({\bf x}_1),\dots,\Pi_{\mathcal{A}_N}({\bf x}_N))$.  For each player $i$, the projection operation $\Pi_{\mathcal{A}_i}({\bf x}_i)$ takes the form 
\begin{equation}
\Pi_{\mathcal{A}_i}({\bf x}_i) = \left(\text{max}\left(0, x_i\left(h\right)-\lambda_i\right), h \in \mathcal{H}\right),\label{pro_form}
\end{equation}
 where $\lambda_i$ is chosen such that the average power constraint is satisfied.  Using (\ref{pro_form}), we rewrite the objective function in (\ref{s_obj}) as
\begin{multline}
\Vert {\bf P} - \Pi_{\mathcal{A}}\left( {\bf P} - F({\bf P})\right)\Vert^2  = \\
\sum_{h \in \mathcal{H},i} \left(P_i\left(h\right)- \text{max}\left\{0, -f_i({\bf P}_{-i};h)-\lambda_i\right\}\right)^2 \\
= \sum_{h \in \mathcal{H},i} \left(\text{min}\left\{P_i(h), \fij{i}{j}{P}+\lambda_i\right\}\right)^2 \\
= \sum_{h \in \mathcal{H},i} \left(\text{min}\left\{P_i(h), P_i(h) + f_i({\bf P}_{-i};h)+\lambda_i\right\}\right)^2 .\label{sim_obj}
\end{multline}
At a NE, the left side of equation (\ref{sim_obj}) is zero and hence each minimum term on the right side of the equation must be zero as well.  This happens, only if 
\begin{equation*}
P_i(h) = \begin{cases}
  0, \text{ if } \fis{i}{j}{P}+\lambda_i > 0, \\
  -\fis{i}{j}{P}-\lambda_i, \text{ otherwise. }
\end{cases}
\end{equation*}
Here, the Lagrange multiplier $\lambda_i$ can be negative, as the projection satisfies the average power constraint with equality.  At a NE Player $i$ will not transmit if the ratio of total interference plus noise to the direct link gain is more than some threshold.\par
\hspace{0.5cm} We now propose a heuristic algorithm to find an optimizer of (\ref{s_obj}).  This algorithm consists of two phases.  In the first phase, it attempts to find a better estimate of a power allocation using the fixed point iteration on the mapping $T({\bf P})$ that is close to a NE.  For $\mathcal{G}_A$ this is algorithm (\ref{f_iter}) itself, which converges to the NE when $\tilde{H}$ is positive semidefinite. When this condition does not hold, then we use it in Algorithm \ref{s_min} to get a good initial point for the steepest descent algorithm of Phase 2.  We will show in Section \ref{ne} that it indeed provides a very good initial point for Phase 2. For games $\mathcal{G}_I$ and $\mathcal{G}_D$ we will provide more justification for Phase 1 by showing that this corresponds to a better response dynamics.  In the second phase, using the estimate obtained from Phase 1 as the initial point, the algorithm runs the steepest descent method to find a NE.  It is possible that the steepest descent algorithm may stop at a local minimum which is not a NE.  This is because of the non-convex nature of the optimization problem.  If the steepest descent method in Phase 2 terminates at a local minimum which is not a NE, we again invoke Phase 1 with this local minimum as the initial point and then go over to Phase 2.  We present the complete algorithm in Algorithm \ref{s_min}.\par
\hspace{0.5cm} In Section \ref{ne} we provide an example when $\tilde{H}$ is positive semidefinite and use the algorithm in \cite{arxiv} to obtain a NE.  We also use Algorithm \ref{s_min} and obtain the same NE (which will be obtained from the first phase itself).  Next we provide examples where $\tilde{H}$ is not positive semidefinite.  Thus the algorithm in \cite{arxiv} may not converge.  The present algorithm provides the NEs in just a few iterations of Phase 1 and Phase 2.
\begin{algorithm}
  \begin{algorithmic}
    \State Fix $\epsilon > 0, \delta > 0$ and a positive integer MAX 
    \State {\bf Phase 1} :  Initialization phase\\
    \State Initialize ${\bf P}_i^{(0)}$ for all $i = 1,\dots,N$.    
    \For {$n = 1 \to \text{MAX} $}
    \State $ {\bf P}^{(n)} = $ $T({\bf P}^{(n-1)})$\vspace{0.2cm}    
    \EndFor
    \State go to Phase 2.\\
    \State {\bf Phase 2} : Optimization phase\\
    \State Initialize $t = 1, {\bf P}^{(t)} = {\bf P}^{MAX}$,
    \Loop
    \State For each $i$, ${\bf P}_i^{(t+1)}$ = Steepest\_Descent($\tilde{{\bf P}}_i^{(t)},i$)
    \State where $\tilde{{\bf P}}_i^{(t)} = ({\bf P}_1^{(t+1)},\dots,{\bf P}_{i-1}^{(t+1)},{\bf P}_i^{(t)},\dots,{\bf P}_N^{(t)})$,
    \State ${\bf P}^{(t+1)} = ({\bf P}_1^{(t+1)},\dots,{\bf P}_N^{(t+1)})$,
    \State $t = t+1$,
    \State Till $\Vert {\bf P}^{(t)} - T({\bf P}^{(t)}) \Vert < \epsilon$
    \If {$\Vert {\bf P}^{(t)} -{\bf P}^{(t+1)}\Vert < \delta $ and $\Vert {\bf P}^{(t)} - T({\bf P}^{(t)})\Vert > \epsilon$}
    \State Go to Phase 1 with ${\bf P}^{(0)} = {\bf P}^{(t)}$
    \EndIf
    \EndLoop
    \Function{$\text{Steepest\_Descent}$}{${\bf P}^{(t)},i$}\\    
    \State $\triangledown f({\bf P}^{(t)}) = (\frac{\partial f({\bf P})}{\partial P_i(h)}\vert_{{\bf P} = {\bf P}^{(t)}}, h \in \mathcal{H})$ 
    \State where $f({\bf P}) = \Vert {\bf P} - T({\bf P}) \Vert^2$ 
    \For {$h \in \mathcal{H}$}    
    \State evaluate $\frac{\partial f({\bf P})}{\partial P_i(h)}\vert_{{\bf P} = {\bf P}^{(t)}}$ using derivative approximation
    \EndFor
    \State ${\bf P}_i^{(t+1)} = \Pi_{\mathcal{A}_i}({\bf P}_i^{(t)} - \gamma_t\triangledown f({\bf P}^{(t)}))$        
    \State return ${\bf P}_i^{(t+1)}$
    \EndFunction    
  \end{algorithmic}  
  \caption{Heuristic algorithm to find a Nash equilibrium}  
  \label{s_min}  
\end{algorithm}
\section{Partial Information games}\label{incomplete}
\hspace{0.5cm} In partial information games, we can not write the problem of finding a NE as an affine variational inequality, because the best response is not water-filling and should be evaluated numerically.  In this section, we show that we can use Algorithm \ref{s_min} to find a NE even for these information games.\par
\subsection{Game $\mathcal{G}_I$}
\hspace{0.5cm} We first consider the game $\mathcal{G}_I$ and find its NE using Algorithm \ref{s_min}.  We follow on similar lines as in Sections \ref{one} and \ref{gen_vi}.  We write the variational inequality formulation of the NE problem.  For user $i$, the optimization at hand is 
\begin{equation}
\text{ maximize } r_i^{(I)}, \text{ subject to } {\bf P}_i \in \mathcal{A}_i,\label{in_pr}
\end{equation}
where $r_i^{(I)} = \sum_{h_i \in \mathcal{I}}\pi(h_i)\mathbb{E}\left[\text{log} \left(1+ \frac{ |h_{ii}|^2 P_i({\bf h}_i)}{1 + \sum_{j \neq i}|h_{ij}|^2P_j({\bf H}_j)}\right)\right]$.  The necessary and sufficient optimality conditions for the convex optimization problem (\ref{in_pr}) are
\begin{equation}
({\bf x}_i - {\bf P}_i^*)^T(-\triangledown_ir_i^{(I)}({\bf P}_i^*,{\bf P}_{-i})) \geq 0, \text{ for all }{\bf x}_i \in \mathcal{A}_i,\label{ns_in}
\end{equation}
where $\triangledown_ir_i^{(I)}({\bf P}_i^*,{\bf P}_{-i})$ is the gradient of $r_i^{(I)}$ with respect to power variables of user $i$.  Then ${\bf P}^*$ is a NE if and only if (\ref{ns_in}) is satisfied for all $i = 1,\dots,N$.
We can write the $N$ inequalities in (\ref{ns_in}) as
\begin{equation}
({\bf x} - {\bf P}^*)^TF({\bf P}^*) \geq 0, \text{ for all }{\bf x} \in \mathcal{A},\label{vi_in}
\end{equation}
where $F({\bf P}) = (-\triangledown_1r_1^{(I)}({\bf P}),\dots,-\triangledown_Nr_N^{(I)}({\bf P}))^T$.  Equation (\ref{vi_in}) is the required variational inequality characterization.  A solution of the variational inequality is a fixed point of the mapping $T_I({\bf P}) = \Pi_{\mathcal{A}}({\bf P} - \tau F({\bf P}))$, for $\tau > 0$.  We use Algorithm \ref{s_min}, to find a fixed point of $T_I({\bf P})$ by replacing $T({\bf P})$ in Algorithm \ref{s_min} with $T_I({\bf P})$.
\subsection{Better response}
\hspace{0.5cm} In this subsection, we interpret $T_I({\bf P})$ as a better response for each user.  For this, consider the optimization problem (\ref{in_pr}).  For this, using the gradient projection method, the update rule for power variables of user $i$ is 
\begin{equation}
{\bf P}_i^{(n+1)} = \Pi_{\mathcal{A}_i}({\bf P}_i^{(n)} + \tau \triangledown_ir_i^{(I)}({\bf P}^{(n)})).\label{btr}
\end{equation}
The gradient projection method ensures that for a given ${\bf P}_{-i}^{(n)}$, $r_i^{(I)}({\bf P}_i^{(n+1)},{\bf P}_{-i}^{(n)}) \geq r_i^{(I)}({\bf P}_i^{(n)},{\bf P}_{-i}^{(n)})$.  Therefore, we can interpret ${\bf P}_i^{(n+1)}$ as a better response to ${\bf P}_{-i}^{(n)}$ than ${\bf P}_i^{(n)}$.  As the feasible space $\mathcal{A} = \Pi_{i=1}^N \mathcal{A}_i$, we can combine the update rules of all players and write 
\begin{equation}
{\bf P}^{(n+1)} = \Pi_{\mathcal{A}}({\bf P}^{(n)} - \tau F({\bf P}^{(n)})) = T_I({\bf P}^{(n)}).
\end{equation}
Thus, the Phase $1$ of Algorithm \ref{s_min} is the iterated better response algorithm.\par
\hspace{0.5cm} Consider a fixed point ${\bf P}^*$ of the better response $T_I({\bf P})$.  Then it implies that, given ${\bf P}_{-i}^*$, ${\bf P}_i^*$ is a local optimum of (\ref{in_pr}) for all $i$.  Since the optimization (\ref{in_pr}) is convex, ${\bf P}_i^*$ is also a global optimum.  Thus given ${\bf P}_{-i}^*$, ${\bf P}_i^*$ is best response for all $i$ and hence NE is also a fixed point of the better response function.  This gives further justification for Phase 1 of Algorithm \ref{s_min}.  We could not provide this justification for $\mathcal{G}_A$ when $\tilde{H}$ is not positive semidefinite.  Indeed we will show in the next section that in such a case Phase 1 often provides a NE for $\mathcal{G}_I$ and $\mathcal{G}_D$ (for which also Phase 1 provides a better response dynamics; see Section \ref{direct} below) but not for $\mathcal{G}_A$.
\subsection{Lower bound}
\hspace{0.5cm} In the computation of NE, each user $i$ is required to know the power profile ${\bf P}_{-i}$ of all other users.  We now give a lower bound on the utility $r_i^{(I)}$ of player $i$ that does not depend on other players' power profiles.\par
\hspace{0.5cm} We can easily prove that the function inside the expectation in $r_i^{(I)}$ is a convex function of ${\bf P}_j(h_j)$ for fixed ${\bf P}_i(h_i)$ using the fact that (\cite{boyd}) a function $f:\mathcal{K} \subseteq \mathbb{R}^n \rightarrow \mathbb{R}$ is convex if and only if $$\frac{d^2 f({\bf x}+t {\bf y})}{dt^2} \geq 0,$$ for all ${\bf x},{\bf y} \in \mathcal{K}$ and $t \in \mathbb{R}$ is such that ${\bf x}+t {\bf y} \in \mathcal{K}$. Then by Jensen's inequality to the inner expectation in $r_i^{(I)}$,
\begin{eqnarray}
r_i^{(I)} & = &\sum_{h_i \in \mathcal{I}}\pi(h_i)\mathbb{E}\left[\text{log} \left(1+ \frac{ |h_{ii}|^2 P_i({\bf h}_i)}{1 + \sum_{j \neq i}|h_{ij}|^2P_j({\bf H}_j)}\right)\right] \nonumber \\
& \geq & \sum_{h_i \in \mathcal{I}}\pi(h_i) \text{log} \left(1+ \frac{ |h_{ii}|^2 P_i({\bf h}_i)}{1 + \sum_{j \neq i}|h_{ij}|^2\mathbb{E}[P_j({\bf H}_j)]}\right)\nonumber \\
& = & \sum_{h_i \in \mathcal{I}}\pi(h_i) \text{log} \left(1+ \frac{ |h_{ii}|^2 P_i({\bf h}_i)}{1 + \sum_{j \neq i}|h_{ij}|^2\overline{P_j}}\right). \label{in_lb}
\end{eqnarray}
The above lower bound $r_{i,LB}^{(I)}({\bf P}_i)$ of $r_i^{(I)}({\bf P}_i,{\bf P}_{-i})$ does not depend on the power profile of players other than $i$.  We can choose a power allocation ${\bf P}_i$ of player $i$ that maximizes $r_{i,LB}^{(I)}({\bf P}_i)$.  It is the water-filling solution given by
\begin{equation*}
P_i(h_i) = \text{ max }\left\{0, \lambda_i - \frac{1 + \sum_{j \neq i}|h_{ij}|^2\overline{P_j}}{|h_{ii}|^2} \right\}.
\end{equation*}
\hspace{0.5cm} Let ${\bf P}^* = ({\bf P}_i^*, {\bf P}_{-i}^*)$ be a NE, and let ${\bf P}_i^{\dagger}$ be the maximizer for the lower bound $r_{i,LB}^{(I)}({\bf P}_i)$.  Then, $r_{i}^{(I)}({\bf P}_i^*,{\bf P}_{-i}^*) \geq r_{i}^{(I)}({\bf P}_i, {\bf P}_{-i}^*) \text{ for all } {\bf P}_i \in \mathcal{A}_i$, in particular for ${\bf P}_i = {\bf P}_i^{\dagger}$.  Thus, $r_{i}^{(I)}({\bf P}_i^*,{\bf P}_{-i}^*) \geq r_{i}^{(I)}({\bf P}_i^{\dagger}, {\bf P}_{-i}^*)$. But, $r_{i}^{(I)}({\bf P}_i^{\dagger}, {\bf P}_{-i}^*) \geq r_{i,LB}^{(I)}({\bf P}_i^{\dagger})$.  Therefore, $r_{i}^{(I)}({\bf P}_i^*,{\bf P}_{-i}^*) \geq r_{i,LB}^{(I)}({\bf P}_i^{\dagger})$.  But, in general it may not hold that $r_{i}^{(I)}({\bf P}_i^*,{\bf P}_{-i}^*) \geq r_{i}^{(I)}({\bf P}_i^{\dagger},{\bf P}_{-i}^{\dagger})$.
\subsection{Game $\mathcal{G}_D$}\label{direct}
\hspace{0.5cm} We now consider the game $\mathcal{G}_D$ where each user $i$ has knowledge of only the corresponding direct link gain $H_{ii}$.  In this case also we can formulate the variational inequality characterization.  The variational inequality becomes
\begin{equation}
({\bf x} - {\bf P}^*)^TF_D({\bf P}^*) \geq 0, \text{ for all }{\bf x} \in \mathcal{A},\label{vi_d}
\end{equation}
where $F_D({\bf P}) = (-\triangledown_1r_1^{(D)}({\bf P}),\dots,-\triangledown_Nr_N^{(D)}({\bf P}))^T$.  We use Algorithm \ref{s_min} to solve the variational inequality (\ref{vi_d}) by finding fixed points of $T_D({\bf P}) = \Pi_{\mathcal{A}}({\bf P} - \tau F_D({\bf P}))$.  Also, one can show that as for $T_I$, $T_D$ provides a better response strategy.  We can also derive a lower bound on $r_i^{(D)}$ using convexity and Jensen's inequality as in (\ref{in_lb}).  The optimal solution for the lower bound is the water-filling solution
\begin{equation*}
  P_i(h_{ii}) = \text{ max }\left\{0, \lambda_i - \frac{1 + \sum_{j \neq i}\mathbb{E}[|H_{ij}|^2]\overline{P_j}}{|h_{ii}|^2} \right\}.
\end{equation*}
\section{Numerical Examples}\label{ne}
\hspace{0.5cm} In this section we compare the sum rate achieved at a Nash equilibrium under the different assumptions on the channel gain knowledge, obtained using the algorithms provided above.  In all the numerical examples, we have chosen $\tau = 0.1$ and the step size in the steepest descent method $\gamma_t = 0.5 \text{ for } t=1$ and is updated after $10$ iterations as $\gamma_{t+10} = \frac{\gamma_{t}}{1+\gamma_{t}}$.  We choose a 3-user interference channel for Examples 1 and 2 below.\par
\hspace{0.5cm} For Example 1, we take $\mathcal{H}_d = \{0.3, 1\}$ and $\mathcal{H}_c = \{0.2, 0.1\}$.  We assume that all elements of $\mathcal{H}_d, \mathcal{H}_c$ occur with equal probability, i.e., with probability 0.5.  Now, the $\tilde{H}$ matrix is positive definite and there exists a unique NE.  Thus, the fixed point iteration (\ref{f_iter}) converge to the unique NE for $\mathcal{G}_A$.  Algorithm \ref{s_min} also converges to this NE not only for $\mathcal{G}_A$ but also for $\mathcal{G}_I$ and $\mathcal{G}_D$.\par
\begin{figure}
  \hspace{-0.5cm}
  \includegraphics[height=5.6cm,width=9cm]{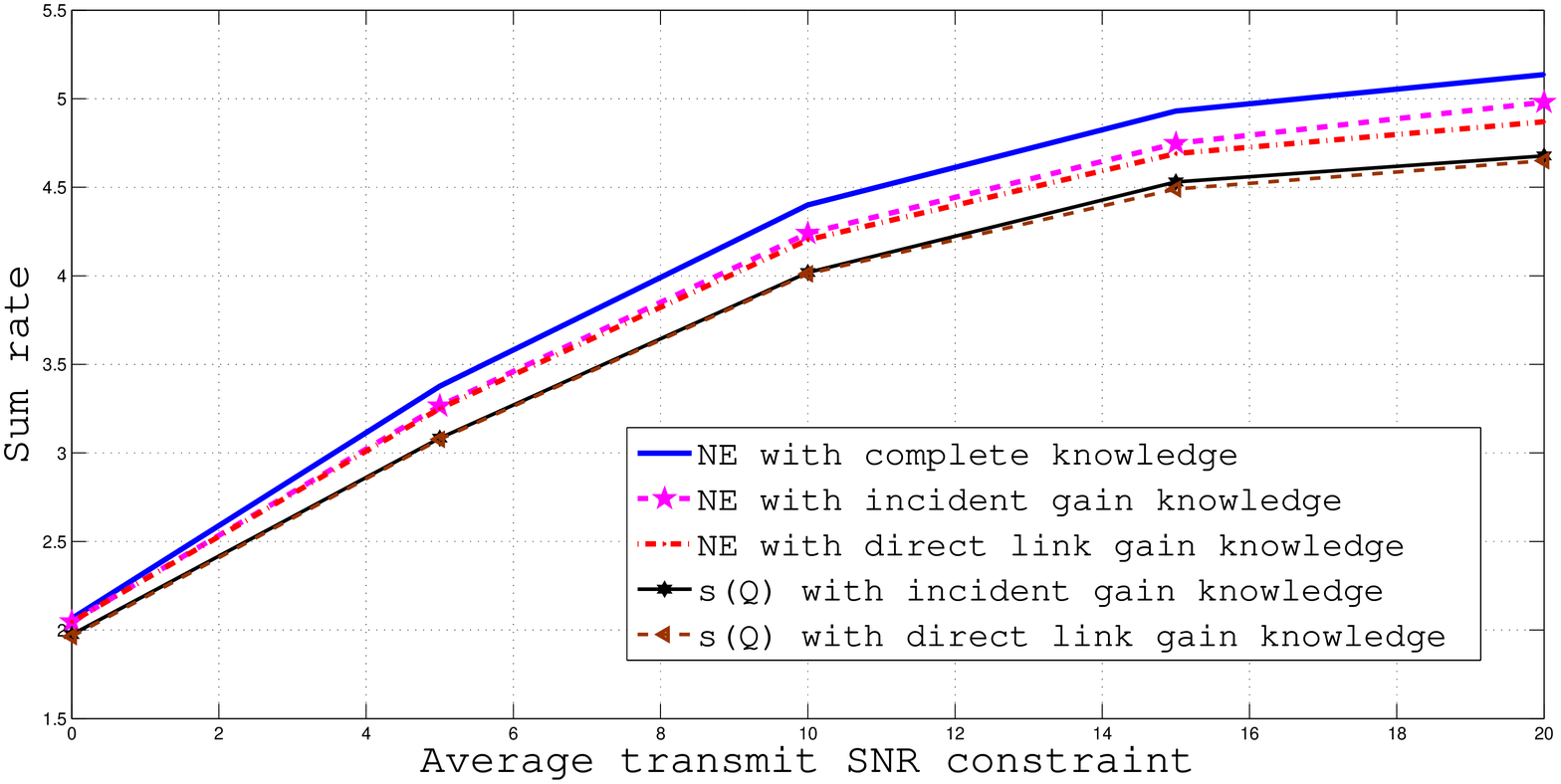}
  \vspace{-0.3cm}
  \caption{Sum rate comparison at Nash equilibrium points for Example 1.}
  \label{plot}    
  \vspace{-0.5cm}
\end{figure}

\hspace{0.5cm} We compare the sum rates for the NE under different assumptions in Figure \ref{plot}.  We have also computed ${\bf Q} = {\bf P}^{\dagger}$ that maximizes the corresponding lower bounds (\ref{in_lb}), evaluated the sum rate $s({\bf Q})$ and compared to the sum rate at a NE.  The sum rates at Nash equilibria for $\mathcal{G}_I$ and $\mathcal{G}_D$ are close.  This is because the values of the cross link channel gains are close and hence knowing the cross link channel gains has less impact.\par
\begin{figure}
  \hspace{-0.5cm}
  \includegraphics[height=5.6cm,width=9cm]{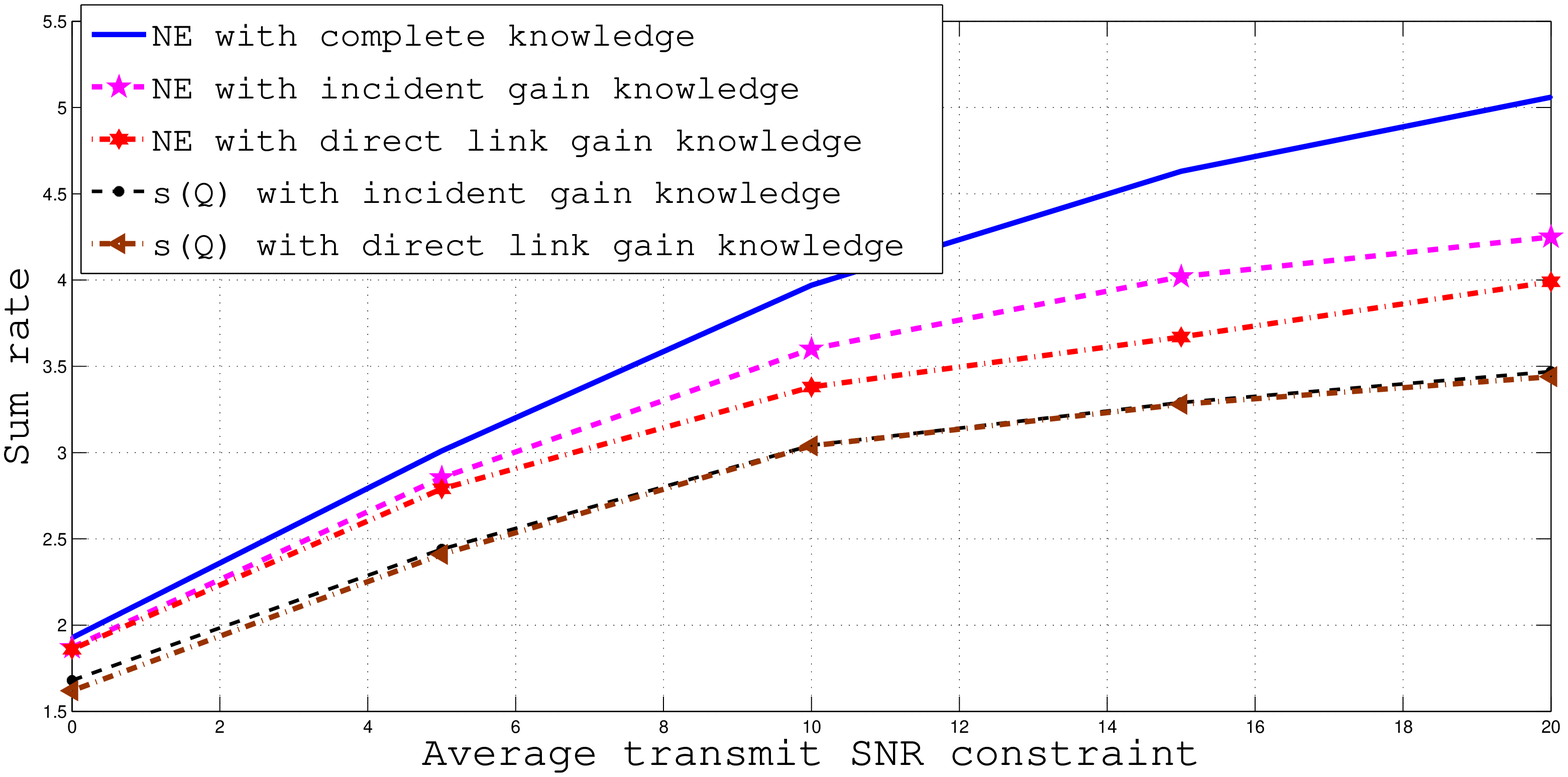}
  \vspace{-0.3cm}
  \caption{Sum rate comparison at Nash equilibrium points for Example 2.}  
  \label{plot2}
  \vspace{-0.5cm}
\end{figure}
\hspace{0.5cm} We now give a couple of examples in which $\tilde{H}$ is not positive semidefinite and hence fixed point iteration (\ref{f_iter}) fails to converge to a NE but Algorithm \ref{s_min} converges to a NE for $\mathcal{G}_A$, $\mathcal{G}_I$ and $\mathcal{G}_D$. \par
\hspace{0.5cm} For Example 2, we take $\mathcal{H}_d = \{0.3, 1\}$ and $\mathcal{H}_c = \{0.1, 0.5\}$.  We assume that all elements of $\mathcal{H}_d, \mathcal{H}_c$ occur with equal probability.  We compare the sum rates for the NE in Figures \ref{plot2}.  Now we see significant differences in the sum rates.\par
\begin{figure}
  \hspace{-0.5cm}
  \includegraphics[height=5.6cm,width=9cm]{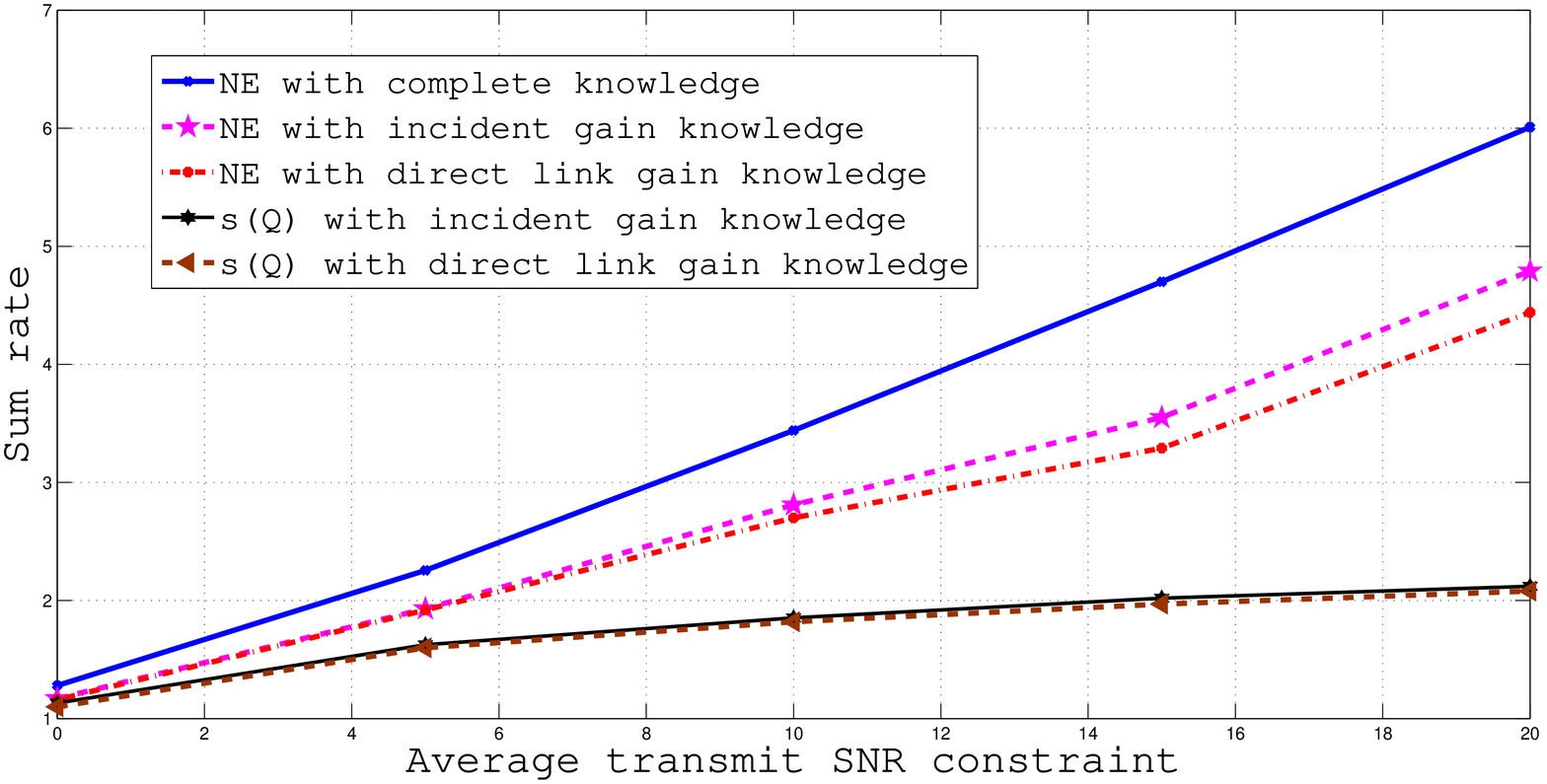}
  \vspace{-0.3cm}
  \caption{Sum rate comparison at Nash equilibrium points for Example 3.}
  \label{plot3}    
  \vspace{-0.5cm}
\end{figure}

\hspace{0.5cm} Consider a 2-user interference channel for Example 3. We take $\mathcal{H}_d = \{0.1, 0.5, 1 \}$ and $\mathcal{H}_c = \{0.25, 0.5, 0.75\}$.  We assume that all elements of $\mathcal{H}_d, \mathcal{H}_c$ occur with equal probability.  In this example also, we use Algorithm \ref{s_min} to find NE for the different cases and the lower bound.  We compare the sum rates for the NE in Figures \ref{plot3}.
\hspace{0.5cm} We further elaborate on the usefulness of Phase 1 in Algorithm \ref{s_min}.  We quantify the closeness of ${\bf P}$ to a NE by $g({\bf P}) = \Vert {\bf P} - T({\bf P}) \Vert$. If ${\bf P}$ is a NE, $g({\bf P}) = 0$ and for two different power allocations ${\bf P}$ and ${\bf Q}$, we say that ${\bf P}$ is closer to a NE than ${\bf Q}$ if $g({\bf P}) < g({\bf Q})$.  We now verify that the fixed point iterations in the initialization phase of Algorithm \ref{s_min} takes us closer to a NE starting from any randomly chosen feasible power allocation.  For this, we have randomly generated $100$ feasible power allocations and run Phase 1 for $MAX = 100$ for each randomly chosen power allocation and compared the values of $g({\bf P})$.  In the following, we compare the mean and standard deviation of the values of $g({\bf P})$ immediately after random generation of feasible power allocations to those after running the initialization phase for the 100 initial points chosen.\par
\hspace{0.5cm} For complete information game, in Example 1: (mean, standard deviation) of values of $g({\bf P})$ after random generation of feasible power allocations at 10dB and 15dB is (230.86, 3.8) and (659.22, 9.21) respectively.  The (mean, standard deviation) for those samples after running the Phase 1 are (0.6260, 0.055) and (2.05, 0.166) respectively at 10dB and 15dB.  Similarly in Example 2: (mean, standard deviation) of $g({\bf P})$ after random generation is (309.12, 4.4) and (950.01, 10.41) respectively at 10dB and 15dB and those after initialization phase are (9.63, 1.83) and (32.26, 6.6).  In Example 3: the mean and standard deviation of $g({\bf P})$ immediately after the random generation are (101.85, 1.1140) at 10dB and (339.97, 9.68) at 15dB.  The (mean, standard deviation) after running the Phase 1 are (0.83, 0.68) at 10dB and (2.82, 1.47) at 15dB.  Thus, we can see that the phase 1 in Algorithm \ref{s_min} provide a much better approximation to a NE than a randomly chosen feasible power allocation.\par
\hspace{0.5cm} For all the three examples, for $\mathcal{G}_I$ and $\mathcal{G}_D$, Phase 1 itself provides the NE.\par
\hspace{0.5cm} We have run Algorithm \ref{s_min} on many more examples and found that it computed the NE, and for $\mathcal{G}_I$ and $\mathcal{G}_D$ the Phase 1 itself provided the NE.
\section{Conclusions}\label{concl}
\hspace{0.5cm} We have considered a channel shared by multiple transmitter-receiver pairs causing interference to each other.  We have modeled this system as a non-cooperative stochastic game.  Different transmitter-receiver pairs may or may not have channel gain information about other pairs' channel gains.  Exploiting variational inequalities, we provide an algorithm that obtains NE in the various examples studied quite efficiently.

\end{document}